\documentclass[aps,prl,reprint, superscriptaddress]{revtex4-1}
\usepackage{blindtext}
\usepackage{graphics}
\usepackage{graphicx}

\usepackage{amssymb}
\usepackage{amsmath}
\usepackage{hyperref}
 \usepackage{color}
\usepackage{comment}

\usepackage{array}
\newcolumntype{L}[1]{>{\raggedright\let\newline\\\arraybackslash\hspace{0pt}}m{#1}}
\newcolumntype{C}[1]{>{\centering\let\newline\\\arraybackslash\hspace{0pt}}m{#1}}
\newcolumntype{R}[1]{>{\raggedleft\let\newline\\\arraybackslash\hspace{0pt}}m{#1}}

\begin{document}

\title{Magnetic-field-induced parity effect in insulating Josephson junction chains}




\author{Timothy Duty}
\email[corresponding author, ]{t.duty@unsw.edu.au.}
	\affiliation{School of Physics, University of New South Wales, Sydney, NSW 2052 Australia}

\author{Karin Cedergren}
	\affiliation{School of Physics, University of New South Wales, Sydney, NSW 2052 Australia}
\author{Sergey Kafanov}
\affiliation{Physics Department, Lancaster University, Lancaster, UK \hspace{2pt} LA1 4YB.} 
\author{Roger Ackroyd} 
	\affiliation{School of Physics, University of New South Wales, Sydney, NSW 2052 Australia}

\author{Jared H. Cole}
	\affiliation{Chemical and Quantum Physics, School of Science, RMIT University, Melbourne,VIC 3001 Australia}

\begin{abstract}

We report the experimental manifestation of even-odd parity effects in the transport characteristics of insulating Josephson junction chains which occur as the superconducting gap is suppressed by applied magnetic fields at millikelvin temperatures. The primary signature is a non-monotonic dependence of the critical voltage, $V_c$, for the onset of charge transport through the chain, with the parity crossover indicated by a maximum of $V_c$ at the parity field $B^*$. We also observe a distinctive change in the transport characteristics across the parity transition, indicative of Cooper-pair dominated transport below $B^*$, giving way to single-electron dominated transport above $B^*$. For fields applied in the plane of the superconducting aluminum films, the parity effect is found to occur at the field, $B^*_{||}$, such that the superconducting gap equals the single-electron charging energy, $\Delta(B^*_{||})=E_C$. On the contrary, the parity effect for perpendicularly applied fields can occur at relatively lower fields, $B^*_\perp\simeq 2\Phi_0/A_I$, depending only on island area, $A_I$. 
Our results suggest a novel explanation for the insulating peak observed in disordered superconducting films and one-dimensional strips patterned from such films.


\end{abstract}

\maketitle


The ground state of a mesoscopic BCS superconductor has been shown to contain an even number of electrons, as long as the single-electron charging energy is less than the superconducting gap, $E_C<\Delta$, and the temperature is less than a characteristic temperature $T^*$. An even-odd parity effect occurs as $T$ exceeds $T^*$, or at very low temperatures, if $\Delta$ becomes lower than $E_C$. This has been demonstrated experimentally as a change from $2e$ to $e$-periodicity in the gate-voltage modulation of superconducting aluminum single-charge transistors and Cooper-pair box qubits\cite{PhysRevLett.69.1993.Averin,PhysRevLett.69.1997.Tuominen,AmarPRL94,LafargePRL93,LafargeNature93}.  A quantitative analysis of the parity effect in hybrid superconductor-semiconductor islands has become an important experimental tool in identifying Majorana modes\cite{albrecht-etal-16}. 
In this Letter, we show that even-odd parity effects strongly affect magnetotransport in insulating Josephson junction chains, and lead to a peak in the critical voltage with magnetic field. Such an insulating peak is observed in homogeneously disordered superconducting films and strips\cite{baturina_etal_2007, vinokur_etal_nature,nguyen_etal_09,ovadia_etal,Schneider18}, which are conjectured to behave as Josephson-coupled grains or droplets\cite{tinkham-book, Fistul_etal_2008}. Josephson junctions chains are also discrete versions of superconducting nanowires

We recently reported results on the thermal parity effect observed in Josephson junction chains very deep in the insulating state, where the characteristic Josephson energy is much less than the Cooper-pair charging energy, $E_\mathrm{J} \ll E_{CP}$, $(E_{CP}=4E_C$)\cite{cedergren-etal-15}. The hallmark of the insulating state---a voltage gap to conduction---was found to vanish sharply at the characteristic temperature $\textrm{k}_\textrm{B}T^* = \Delta / \ln N_\mathrm{eff} \simeq \Delta/9$, which coincides with the presence of $\sim1$ thermally excited BCS quasiparticle per island, where $N_\mathrm{eff} (T)\approx \mathcal{V}\rho(0)\sqrt{2\pi k_\mathrm{B}T\Delta(T)}$ is the effective number of states arising from integration over the quasiparticle density of states, $\mathcal{V}$ is the volume of the island and $\rho(0)$ is the density of states for the normal metal at the Fermi energy\cite{PhysRevLett.69.1993.Averin}. 

In a more recent Letter, we showed that insulating Josephson junction chains in zero magnetic field, with $E_\mathrm{J}\,\sim\,E_{CP}$, behave as 1D Luttinger liquids, pinned by offset charge disorder, and therefore can be understood as a circuit implementation of the one-dimensional Bose glass\cite{cedergren-etal-17}. The key result was that the voltage gap, $V_c,$ for the onset of conduction arises from collective depinning of Cooper-pair quasicharge, and is inversely related to the localisation length as calculated by Giamarchi and Schultz\cite{giamarchi-schultz-88}. We also found $V_c$ to be proportional to the number of junctions in the chain, $N$, and decreasing as a power law in bandwidth, $W$. The Bloch bandwidth $W$ is prescribed by the single-junction theory of quasicharge energy bands\cite{likharev-zorin,wilkinson_etal_18}, and can be envisioned as the amplitude for coherent tunneling of flux quanta. $W$ decreases exponentially with $\sqrt{8E_\mathrm{J}/E_C}$.


 In this work, we examine the dependence of $V_c$ on both parallel and perpendicularly-applied magnetic fields, finding a non-monotonic dependence of the critical voltage with magnetic field. Below an orientation-dependent cross over field, $B^*$, the critical voltage $V_c$ increases with field in accordance with increasing $W$, which occurs by suppression of the superconducting gap, $\Delta$, and hence decreasing $E_\mathrm{J}$. Above $B^*$, however, the critical voltage {\em decreases} with field, until finally becoming independent of magnetic field above the superconducting critical field $B_c$. 
 
 We show that the peak behavior in the critical voltage, along with the change in current-voltage characteristics (IVC's) at $B^*$ reveal a parity crossover where collective depinning of Cooper pairs gives way to that of single-electron charges. Moreover, when the field is applied perpendicular to the island films, the parity effect can be accompanied by a change in vorticity of the superconducting islands. In other words, the parity effect occurs in sync with formation of the single-vortex state. Charge transport above $B^*$ for both geometries is found to be markedly different from that below, lending additional support to our interpretation of the insulating peak as an experimental signature of an even-odd parity transition.
 
The interplay of parity and vortex dynamics has been discussed theoretically by Khaymovich \textit{et al.}\cite{khaymovich-etal-2015} for single-island devices in the context of charge pumping, and single-vortex trapping was observed experimentally in hybrid normal-superconducting-normal transistors\cite{taupin_etal_16}.

Although the parity effect in superconducting single-charge transistors and Cooper-pair boxes has been well known for some years, it has received very little attention in studies of Josephson junction arrays. A theoretical paper by Feigel'man \textit{et al.}\cite{FeigelmanJETP} pointed out the implications of the parity effect on the experimental search for the Kosterliz-Thouless charge-unbinding transition. The parity effect for chains in the sequential tunneling limit, $E_\mathrm{J} \ll E_C$, was studied theoretically by \cite{Cole2015}, inspired by the experimental results of Bylander \textit{et al.}\cite{PhysRevB.76.020506R.Bylander}. Detailed treatments of parity effects in junction arrays with $E_\mathrm{J}$$
\,$$\sim$$\,$$E_C$,  two-dimensional disordered superconducting films, and one-dimensional superconducting nanowires, however, are conspicuously absent. 

\begin{table}
\begin{ruledtabular}
\caption{Devices} 

\begin{tabular}{ C{0.22\linewidth}  C{0.09\linewidth}  C{0.09\linewidth}  C{0.11\linewidth}  C{0.07\linewidth} C{0.09\linewidth} C{0.09\linewidth} C{0.12\linewidth} }

Device	& $E_C$ ($\mu$eV) & $E_\mathrm{J0}$ ($\mu$eV)& $g$ & $\vec{B}$ & $B^*$ (mT) & $B_0$ (mT) & $A_{J}$ ($\mu$m)$^2$\\ 
\hline \\
	AS7, N=250    &95.2  & 87.9 &0.23& $\parallel$ & 310 & 455 & 0.0110 \\
	BS1, N=250   &114 & 60.2&0.13& $\perp$& 58.5 & 75.6 & 0.0094\\
	BS2, N=250  & 114  & 74.7  &0.16& $\perp$ & 58.3 & 80.9 & 0.0076 \\
	BS3, N=250   &101  & 92.0& 0.23& $\perp$ & 60.9 & 89.0 & 0.0091 \\
	BS4, N=250   &94.0  & 104& 0.28& $\perp$ & 59.6 & 90.7 & 0.0106   \\
	BS6, N=250   &82.8  & 129&0.39& $\perp$ & 56.9 & 90.2 &0.0117  \\
	BS7, N=250    &71.5  & 167&0.58& $\perp$ & 54.5 & 85.0 & 0.0123  \\
	CS3, N=250   &81.6  & 151 &0.46& $\parallel$ & 332 & 455 & 0.0121  \\
	CS5, N=250   &64.8 & 230 &0.89& $\parallel$ &  346 & 434 & 0.0145 \\
	CS6, N=250   &60.4  & 275 &1.13& $\parallel$ & 355 & 426 & 0.0161   \\
	DS1, N=250    &68.0  & 223  &0.82& $\perp$ & 53.2 & 93.7 & 0.0142 \\
	LS1, N=100    &54.7  & 45.2& 0.21 & $\parallel$ & 362 & 401 &- \\
	LS2, N=200   &57.5 & 44.1& 0.19  & $\parallel$ & 364  & 417 &- \\

\end{tabular}
\end{ruledtabular}
\end{table}

\begin{figure}

\includegraphics[width=0.92\columnwidth]{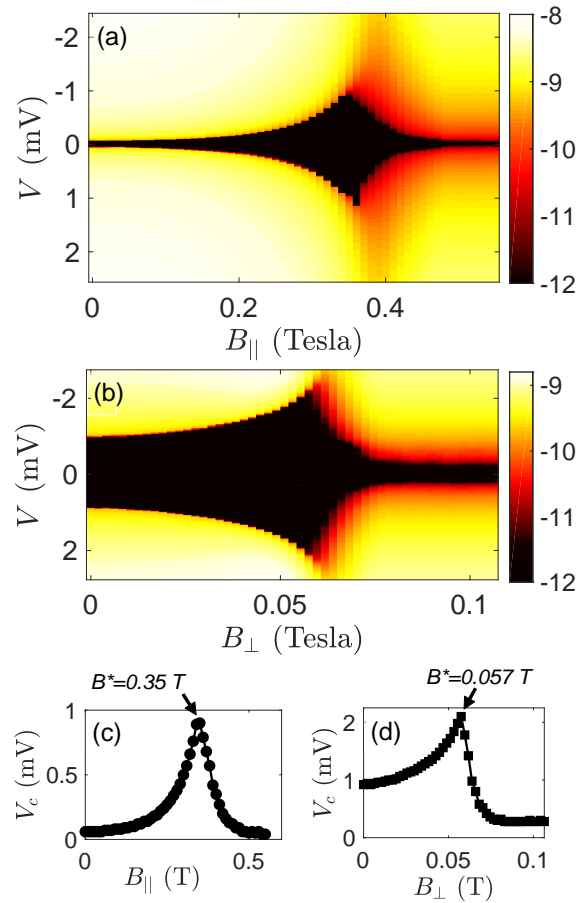}

\caption{Measured current on a logarithmic scale, ($\log_{10}|I|$), versus bias voltage $V$ and magnetic field for (a) device C5 in a parallel magnetic field $B_{||}$, and (b) for device B6 in a perpendicular field $B_{\perp}$.  Critical voltage $V_c(B_{||}$ found for device C5 (c), and $V_c(B_\perp$ for device B6 (d).   The parity transition occurs at $B^*$, the field where critical voltage, $V_c$, is maximum. The critical field $B_c$ is identified as the field at which $V_c$ becomes independent of field.}

\end{figure}


We have fabricated and measured a large ensemble of Al/AlO$_ {x}$/Al single-junction chains\cite{cedergren-etal-17}. Several families of devices were produced, where within each family, we varied the junction area, $A_J$, in order to geometrically tune the ratio $g=E_\mathrm{J0}/4E_C$, where $E_\mathrm{J0}$ is the Josephson tunneling energy at zero magnetic field (see  SM\cite{supplementary}). For each device, we first obtain an accurate measure of the average junction charging energy, $E_C$, from the voltage offset, $V_\mathrm{off}$, of each device found from extrapolating its linear current-voltage characteristic (IVC) from large voltage biases, $V \gg 2 N \Delta_0/e$ , where $\Delta_0$ is the superconducting gap in zero magnetic field, and $N$ is the number of junctions in the chain. As noted in \cite{cedergren-etal-15,supplementary,Tighe93}, the experimentally determined charging energy is found as$\text{,}$ $E_C = e V_\mathrm{off}/N$, and the junction Josephson energy $E_\mathrm{J0}$ across the chain is found from the normal state conductance using the Ambegaokar-Baratoff relation,  $E_\mathrm{J0}=\Delta_0R_Q/2R_N$, where $R_Q$ is the superconducting quantum of resistance, and $R_N$ is the junction resistance in the normal state The experimentally determined device parameters are listed in Table I.

Next we measured the critical voltage $V_\mathrm{c}$, deep in the subgap region, $V \ll 2 N \Delta_0/e$. In addition to the zero-field measurements, IVC's for some devices were also measured with varying parallel, or perpendicularly-applied magnetic fields. Most of the measured devices have non-hysteretic IVC's in this region, for all values of applied field, however,  a few devices exhibited hysteretic IVC's for some values of magnetic fields. In this work, we are interested in using the voltage gap as a probe of the parity effect. We therefore take $V_c$ as the return voltage, that is, the voltage magnitude at which the device returns to the zero-current state (current less than the noise floor), when stepping from the non-zero current state.  The return voltage is found in all cases to be characterized by an extremely narrow distribution (smaller than the experimental resolution). We note that unlike the situation for nanowires and films based on disordered superconducting films\cite{Schneider18}, the voltage gaps we observe are very hard:  we do not find an observable zero-bias resistance arising from a postulated parallel quasiparticle conductance channel, and therefore we have no need to subtract a finite current, $V/R_N$, in order to observe a robust critical voltage. This indicates that our junction chains are significantly more homogeneous than devices based on disordered films.




\begin{figure}
	\includegraphics[width=0.97\columnwidth]{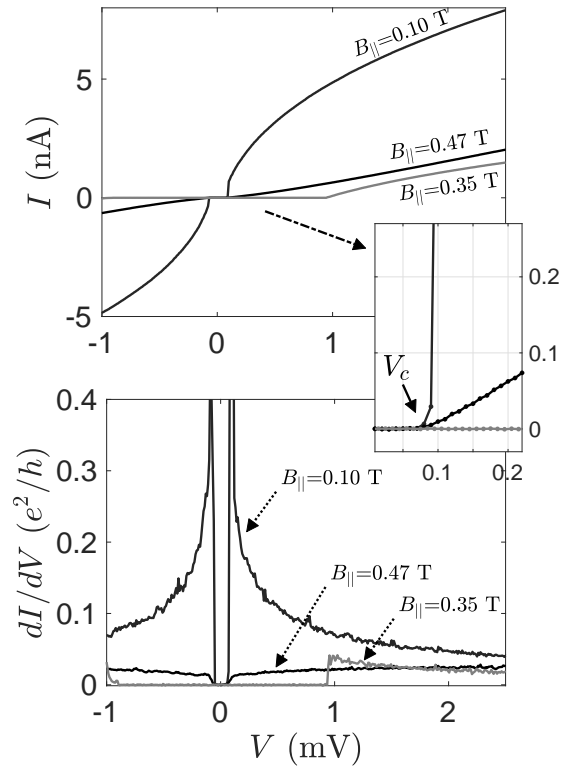}
\caption{Transport data $I$\,\textit{vs.}\,$V$ (upper plot), and  $dI/dV$ \,\textit{vs.}\,$V$  (lower plot) for device C5 at selected values of parallel magnetic field: below the parity field, $B_{||}$\,=\,$0.1$\,T, at the parity field, $B^*_{||}=0.35$ (grey), and above the parity field, $B_{||}=0.47$\,T. Although $V_c$ is approximately equal at 0.1 T and 0.47 T (inset), the transport characteristics are found to be substantially different. The conductance above $V_c$ decreases with $V$ for fields below $B^*$ as expected for Cooper-pair dominated transport, and increases with $V$ for $B_{||}$ above $B^*$, indicative of single-electron transport. Similar behavior is observed for perpendicularly-applied fields (see SM).}

\end{figure}

Figure 1, (a) and (b), show logarithmic scale ($\log_{10} |I|$) image plots of the IVC's for stepped magnetic fields for devices C5 and B6. Device C5 was measured in a parallel magnetic field, while device B6 was measured in a perpendicular field. Both devices were fabricated during the same fabrication run in neighboring squares on the same chip. The critical voltage for each value of magnetic field is identified as the voltage magnitude for which the measured current becomes less that the noise floor for the measurement, which is $\lesssim 1$ pA. One can readily identify a peak in the critical voltages, as shown in Fig. 1, (c) and (d). We estimate the precise field for the peak, $B^*_{||}$ (or $B^*_\perp$), by fitting to few points of  the experimentally determined $V_c$ around it's maximum.

From Figure 2, one notes that the transport data for both parallel and perpendicular fields are remarkably different above and below their respective parity fields. This can be seen more directly in the individual IVC's, as shown for example for Fig. 2, where we plot data for device C5 at magnetic fields both above and below $B^*_{||}$, fields where $V_c$ is approximately equal. We note that all measured devices show qualitatively different transport characteristics above compared to below their $B^*$'s (see Supplemental Material\cite{supplementary} for more examples).

In Figure 2 (lower plot), it is evident that for $B_{||}$\,=\,$0.1$\,T ($< B^*$), the conductance $dI/dV$ is strongly peaked just above $\pm V_c$, and strictly decreasing with $|V|$ outside the voltage gap. This supercurrent-like feature is indicative of Cooper-pair dominated transport. Conversely, for $B_{||}=0.47$\,T ($> B^*$), the conductance increases monotonically outside of the voltage gap. We note that both sets of data asymptotically approach each other. This can be understood qualitatively, as below $B^*$, increasing charge transport invariably populates higher bands in quasicharge \textit{via} Landau-Zener tunneling. The ensuing relaxation from higher bands produces BCS quasiparticles, which eventually suppress the even-odd free energy difference that permits Cooper-pair tunneling to dominate.

\begin{figure}
	\includegraphics[width=0.98\columnwidth]{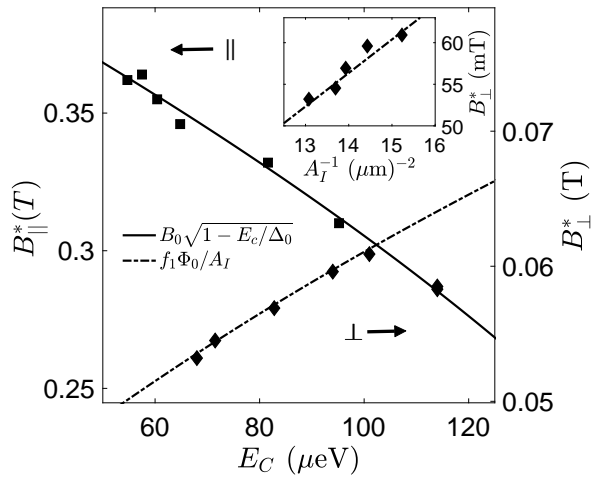}
\caption{Measured parity field $B^*$ \textit{versus} experimentally determined single-electron charging energy, $E_C$, for devices measured in parallel (squares, left axis), and perpendicular (diamonds,  right axis), magnetic fields. The solid  black line follows from suppression of the homogeneous superconducting gap, $\Delta(B)$\,=\,$\Delta_0[1-(B/B_0)^2]$, and setting, $\Delta(B^*)$\,=\,$E_C$, to find $B^*$. The dashed line is a fit to $B^*$\,=\,$f\Phi_0/A_I$ (using $E_C$ as a  proxy for $1/A_I$), where $A_I$\,=\,$wl$, $w$ is the island film width, and $l$, the between-junction island length. A fit to perpendicular field devices having $E_C<110\,\mu$eV, yields $f_1$\,=\,$2.00\pm0.05$. The inset shows $B^*_\perp$ \textit{vs.} $A_I^{-1}$ for these devices and the fit directly.
}

\end{figure}

In Figure 3, we plot the experimentally determined parity fields for six devices in parallel magnetic fields (squares, left axis), and seven devices in perpendicular magnetic fields (diamonds, right axis), as a function of their experimentally determined single-electron charging energy. First we focus on the parallel field data. The solid line follows from suppression of the superconducting gap according to $\Delta=\Delta_0\left(1-B^2/B_0^2\right)$, which we previously found adequate to describe suppression of the superconducting gap in comparably-sized aluminum islands in a parallel magnetic fields\cite{cedergren-etal-15}. Setting $E_C=\Delta$, at $B^*$, one finds $B^*=B_0\sqrt{1-E_C/\Delta_0}$. $B_0$ is known as the pair-breaking parameter, which is expected to be of the same order as the critical field $B_c$. The solid line represents the best fit to the parallel field data, resulting in $B_0=0.42$\,T, which is comparable to the average critical field found for these devices, $B_c=0.48$\,T. 

An alternative method to estimate the depairing parameter, is to fit $V_c(B)$ at fields below $B^*$ to the scaling law behaviour of $V_c(W)$, as detailed in Ref.\cite{cedergren-etal-17}. Since $V_c$ depends on $E_\mathrm{J}$ through $W$, the dependence of $V_c$ on $\Delta$ and hence $B$ can be found assuming $E_\mathrm{J}$ to be given by the Ambegaokar-Baratoff formula, resulting a field dependence $E_\mathrm{J}(B)=E_\mathrm{J0}\left(1-B^2/B_0^2\right)$ (see Supplemental Material\cite{supplementary}). Averaging over the parallel field devices (see Table I), we find $B_0=0.43\pm0.02$\,T, which agrees extremely well with $B_0=0.42$\,T found above. 

Considering now the devices measured in a perpendicular field, it is clear that $B^*_\perp(E_C)$ does not follow the dependence, $B_0\sqrt{1-E_C/\Delta_0}$, in particular, for the five devices with $E_C<110\, \mu$eV. We can, however, estimate the depairing parameter, $B_0$, as discussed in the preceding paragraph. 
Averaging over the perpendicular field devices, we arrive at $B_0=0.086\pm0.003$T, which agrees with the observed critical perpendicular field, $B_c = 0.086$T. For perpendicular field devices, the solid line of Figure 3 corresponds to $B_0=0.086$\,T. The experimental dependence of $B^*_\perp$ on $E_C$ for devices with $E_C \lesssim 110$\,$\mu$eV is nearly orthogonal to that predicted by the black line. This suggests a different mechanism driving the even-odd parity effect: in a perpendicular field, the parity effect can occur in sync with trapping of a single flux quanta. Concomitant with formation of the vortex state,  the average superconducting gap across the island becomes nearly zero, destroying the free energy difference between even and odd parity.

Extensive studies, both theoretical and experimental, have been reported detailing the vortex states of mesoscopic superconducting islands, typically within the context of finding eigenvalues of the linearised Ginzburg-Landau equations\cite{buisson_etal_90,geim_etal_97,benoist-zwerger-97,schweigert_peeters_98,chibotaru_etal_00, chibotaru_etal_05}. For a thin superconducting disc of radius, $R$, it is found, \cite{benoist-zwerger-97,schweigert_peeters_98}, that the single vortex state becomes energetically favourable for an applied flux $\Phi\simeq f_1$\,$\Phi_0$, or $B=f_1$\,$\Phi_0/\pi R^2$, with $f_1$\,$=$\,$1.924$. For a square, the transition is found to occur at $f_1\simeq2.0$\cite{chibotaru_etal_00}, 

Our chains are composed of thin films structured approximately as rectangular islands having a length $l\simeq 890$ nm, thickness $d$\,$=$\,$30$ nm, and varying width (we varied the junction area across each family to modulate the ratio, $g=E_\mathrm{J}/E_C$, while using the same oxidation parameters\cite{cedergren-etal-17}). 
We have analysed SEM images of our devices finding device-averaged junction widths, $w$, ranging from 80 to 120 nm. We find that in accordance with a parallel plate capacitor model of the junctions, the charging energy indeed scales inversely with device-averaged junction area, $A_J=w^2$. For the family of devices measured in perpendicular field we find, $E_C=0.96/A_J$, with $A_J$ in $\mu$m$^2$\cite{ supplementary}. This also gives us a relation between charging energy and average island area, since the island widths equal those of the junctions, $A_I=w (0.89-2w)$ $\mu$m$^2$.

If we neglect the region of the islands making up the junction, \textit{i.e.} we take for the island area, $A_I$,  the area between the junctions, remarkably, we find that we can fit the larger area devices ($E_C \lesssim 110$ $\mu$eV), measured in perpendicular fields, to a single curve, $B^*_\perp=f_1 \Phi_0/A_I$, with$f_1=2.00\pm0.05$, as show by the dashed lines in Figure 3 (main plot and inset). 

In the dirty limit, which applies to our films, the coherence length is given by $\xi_0=\sqrt{\hbar D/\Delta_0}$, where $D$ is the diffusion coefficient, which can be deduced from the conductivity, $\sigma$, using Einstein's relation, $\sigma=N(0)e^2D$, where $N(0)$\,=\,$2.15 \times 10^{47}$\,J$^{-1}$m$^{-3}$ is the density of states at the Fermi level for aluminum, and $e$ is the electron charge. From the measured resistivity of our 30 nm films, we estimate a coherence length, $\xi_0\sim 60$\,nm. The critical field for a large film, $B_{c2}=\Phi_0/2\pi\xi_0^2$, for $\xi_0\sim 60$\,nm gives an estimate $B_{c2}\simeq 90$\,mT, which roughly agrees with the observed critical fields of our devices in perpendicular fields.  

For thin films in parallel applied fields, calculations based on Ginzburg-Landau theory find that the vortex state can only occur when the thickness, $d$, is greater than $1.84\,\xi_0$\cite{fink69,tinkham-book}. This corresponds well to our parallel field results. Since $d<\xi_0$, we see only homogenous suppression of the superconducting gap up to the parity field where $\Delta(B^*) = E_C$. For perpendicular fields, analyses of the vortex states of mesoscopic islands based on the linearized Ginzburg-Landau equations show that the single vortex state is stable only for film widths greater than a critical value $w_c \sim 2\xi_0$\cite{schweigert_peeters_98,chibotaru_etal_00}. The measured widths of our devices appear to cross this borderline width. For such strong confinement, ($w\gtrsim\xi_0$), however, an analysis based on the nonlinear Ginzburg-Landau equations, or the Bogoliubov-de Gennes equations, may be required for a more quantitative comparison.

In conclusion, we find that the non-monotonic dependence and peak in the voltage gap of insulating Josephson junction chains with magnetic field arises from an even-odd parity effect. Below the parity field, the ground state of our insulating junctions chains is that of a one-dimensional Bose glass of localized Cooper-pairs, with the onset of transport arising from depinning of the compressible Cooper-pair quasicharge\cite{cedergren-etal-17}. Above $B^*$, odd parity quasicharge bands become accessible, so that the nature of the ground state becomes that of a Fermi glass, and depinning involves single-electron charges rather than Cooper pairs. Moreover, in a perpendicular field the transition can occur simultaneously with trapping of a single flux quantum in the thin-film superconducting islands. 

Our results are relevant for the insulating peak observed in disordered superconducting films, and strips patterned from such films, which are postulated to form a network of Josephson-coupled superconducting droplets\cite{baturina_etal_2007, vinokur_etal_nature,nguyen_etal_09,ovadia_etal,Schneider18}. A current explanation for such data is the formation of random SQUID loops in the network\cite{Fistul_etal_2008}. Our results suggest such a peak arises from an even-odd parity effect, and may occur with formation of the vortex state in the effective superconducting droplets. We suggest that even-odd parity effects could also be observed in superconducting nanowires.

TD acknowledges useful discussions with Alexander Shnirman. This work was supported by the ARC Centre of Excellence for Engineered Quantum Systems, CE11000101. Devices were fabricated at the UNSW Node of the Australian National Fabrication Facility. JHC is supported by the Australian Government's NCI National Facility through the National Computational Merit Allocation Scheme, and the ARC Centre of Excellence in Future Low-Energy Electronics Technologies (FLEET), CE170100039.

\renewcommand{\thefigure}{S\arabic{figure}}
\setcounter{figure}{0}   
\newpage
	

\begin{center}
\large  \textbf{Supplemental material to:  ``Magnetic-field-induced parity effect in insulating Josephson junction chains"}
\end{center}

\section{Devices}
Al/AlO$_{x}$/Al junction chains were fabricated by standard e-beam lithography followed by two angle evaporation of aluminum with \textit{in situ} oxidation between the two evaporation steps. Each film is 30\,nm thick. Devices were fabricated on silicon substrates (n-doped) with approximately 300\,nm thermally grown SiO\textsubscript {2} on top. Six devices  were measured in magnetic fields parallel to the substrate, and 7 devices in magnetic fields perpendicular to the substrate. The experimentally determined and parameters for the devices are listed in Table I.

The chain average single-electron, single-junction charging energy, $E_C=e^2/2C_J$, where $C_J$ is  the average junction capacitance, is extracted from the voltage offset, $V_\mathrm{off}$, of each device, found from extrapolating its linear IVC from large voltage bias. This is a standard procedure which has been used previously by many groups, and is described in \cite{cedergren-etal-17,cedergren-etal-15, tighe-etal-93, kuzmin-etal-92, delsing-92}. The experimentally determined charging energy is found as, $E_{C}= e V_\mathrm{off}/N$, where $N$ is the number of junctions in the chain. One must take care that $V_\mathrm{off}$ is extrapolated from sufficiently high voltages above the Coulomb blockade where the conductance $d I/d V$\ is measured to be constant, in order to get an unbiased estimate of $E_C$. 

The average junction Josephson energy in zero field, $E_\mathrm{J0}$, is determined from the linear conductance at large voltage bias, $eV \gg 2N\Delta$, using the Ambegaokar-Baratoff relation $E_\mathrm{J0} =\frac{\Delta_0}{2} \frac{R_\mathrm{Q}}{R_\mathrm{J}}$, where $R_\mathrm{Q}$ is the superconducting resistance quantum, and $R_\mathrm{J}$ the inverse of the large-bias linear conductance $G_{\Omega }$ of the array divided by $N$. Here $\Delta $ is taken to be 210\,$\mu$eV, as found in \cite{cedergren-etal-15}. 

The experimentally determined parity field is denoted $B^*$, and the depairing parameter, $B_0$, as determined from a fit to the critical voltage $V_c(B)$ for $B<B^*$, assuming $E_\mathrm{J}=\Delta(B) R_Q G_{\Omega }/2$, with $\Delta(B)=\Delta_0(1-B^2/B_0^2)$, using the scaling of $V_c$ with Bloch bandwidth $W$ (which depends on $E_\mathrm{J}$ and $E_C$), detailed in Ref. \cite{cedergren-etal-17}, described in more detail below.

\begin{table}
\begin{ruledtabular}
\caption{Devices} 

\begin{tabular}{ C{0.22\linewidth}  C{0.09\linewidth}  C{0.09\linewidth}  C{0.11\linewidth}  C{0.07\linewidth} C{0.09\linewidth} C{0.09\linewidth} C{0.12\linewidth} }

Device	& $E_C$ ($\mu$eV) & $E_\mathrm{J0}$ ($\mu$eV)& $g$ & $\vec{B}$ & $B^*$ (mT) & $B_0$ (mT) & $A_{J}$ ($\mu$m)$^2$\\ 
\hline \\
	AS7, N=250    &95.2  & 87.9 &0.23& $\parallel$ & 310 & 455 & 0.0110 \\
	BS1, N=250   &114 & 60.2&0.13& $\perp$& 58.5 & 75.6 & 0.0094\\
	BS2, N=250  & 114  & 74.7  &0.16& $\perp$ & 58.3 & 80.9 & 0.0076 \\
	BS3, N=250   &101  & 92.0& 0.23& $\perp$ & 60.9 & 89.0 & 0.0091 \\
	BS4, N=250   &94.0  & 104& 0.28& $\perp$ & 59.6 & 90.7 & 0.0106   \\
	BS6, N=250   &82.8  & 129&0.39& $\perp$ & 56.9 & 90.2 &0.0117  \\
	BS7, N=250    &71.5  & 167&0.58& $\perp$ & 54.5 & 85.0 & 0.0123  \\
	CS3, N=250   &81.6  & 151 &0.46& $\parallel$ & 332 & 455 & 0.0121  \\
	CS5, N=250   &64.8 & 230 &0.89& $\parallel$ &  346 & 434 & 0.0145 \\
	CS6, N=250   &60.4  & 275 &1.13& $\parallel$ & 355 & 426 & 0.0161   \\
	DS1, N=250    &68.0  & 223  &0.82& $\perp$ & 53.2 & 93.7 & 0.0142 \\
	LS1, N=100    &54.7  & 45.2& 0.21 & $\parallel$ & 362 & 401 &- \\
	LS2, N=200   &57.5 & 44.1& 0.19  & $\parallel$ & 364  & 417 &- \\

\end{tabular}
\end{ruledtabular}
\end{table}

\section{\textbf{Critical voltage}}
Following the characterisation of the Coulomb and Josephson energy scales for each device, we measure the critical voltage $V_\mathrm{c}$, at biases deep in the subgap region, $eV \ll 2N\Delta $, for varying magnetic fields. Devices with low values of $g$ typically have non-hysteretic IVC's in this region. At larger values of $g$, some devices exhibit hysteretic IVC's for some values of magnetic field.  In this work, we are interested in using the voltage gap as a probe of the parity effect. We therefore take $V_c$ as the magnitude of voltage at which the device returns to the zero-current state (current less than the noise floor), when stepping from the non-zero current state, since the return voltage is found in all cases to be characterized by an extremely narrow distribution (smaller than the experimental resolution). 

\section{Junction area determines charging energy}

As described in Ref. \cite{cedergren-etal-17}, we varied the junction area across different chains, in order to vary the coupling constant $g$. The junctions of our device can be considered as parallel plate capacitors, that is, the junction capacitance, $C_J$, is proportional to the junction area, $A_J=w^2$, where $w$ is the width of the junctions (and islands) making up the device. We have analysed scanning electron microscope (SEM) images of most of the devices we have measured to extract device-averaged junction widths, and correlated the average junction areas with measurement of $V_\mathrm{off}/N$, which provides an experimental estimate of $E_c$, or equivalently, $C_J$. In Figure S1, we plot both the experimentally determined $C_J$ \textit{vs.} junction area, $A_J$  (top plot), or alternatively, $E_C$  \textit{vs.}  the inverse of the area, $A_J^{-1}$   (bottom plot). We find clear linear relations between these quantities, which confirms the validity of the parallel plate capacitance model. For the purpose of comparing data and theory, this also allows us to express the device-averaged junction (and island) width, $w$, from the measured charging energy, $E_C$, as $w^{-2}=1.04\times E_C$.

\begin{figure}
	\includegraphics[ width=\columnwidth]{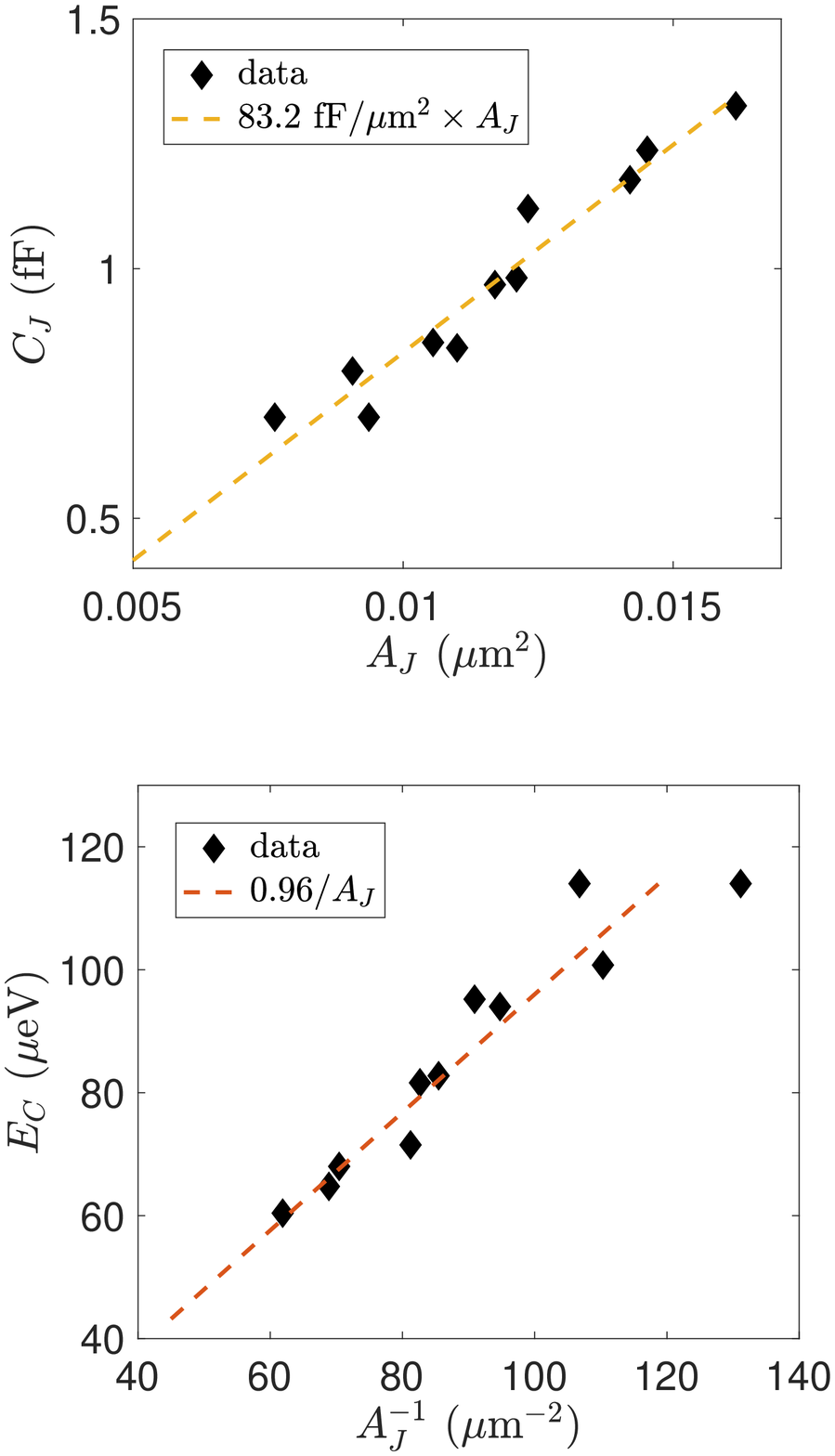} 
	\caption{(Upper plot) Experimentally determined $C_J$ as found from $C_J=e^2N/V_\mathrm{off}$, as a function of junction areas, $A$, extracted from SEM images.  (Lower plot) The same data, but shown as $E_C=eV_\mathrm{off}/N$ \textit{versus} $A_J^{-1}$.}
	 
\end{figure}

\begin{figure}
	\includegraphics[ width=\columnwidth]{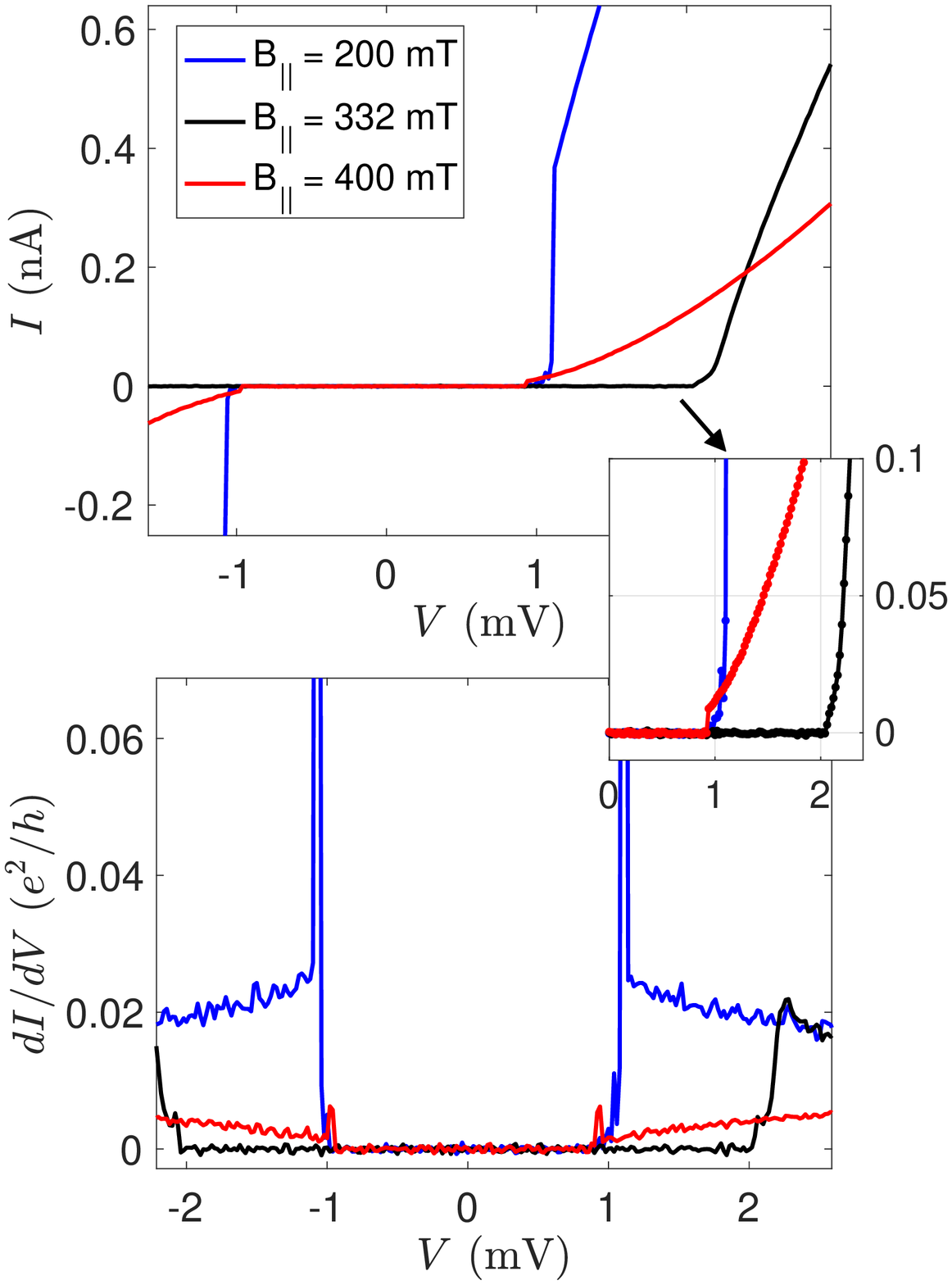} 
	\caption{Transport data $I$ \textit{vs.} $V$ (upper plot), and  $dI/dV$  \textit{vs.} $V$  (lower plot) for device C3 at selected values of parallel magnetic field: below the parity field, $B_{||}$\,=\,$200$\,mT (blue), at the parity field, $B^*_{||}=332$\,mT (black), and above the parity field, $B_{||}=400$\,mT (red).  Although $V_c$ is approximately equal for blue and red data (inset), the transport characteristics are found to be substantially different. The conductance above $V_c$ decreases with $V$ for fields below $B^*$, as expected for Cooper-pair dominated transport, and increases with $V$ for $B_{||}$ above $B^*$, indicative of single-electron transport. Similar behavior is observed for perpendicularly-applied fields}	 
\end{figure}

\begin{figure}
	\includegraphics[ width=\columnwidth]{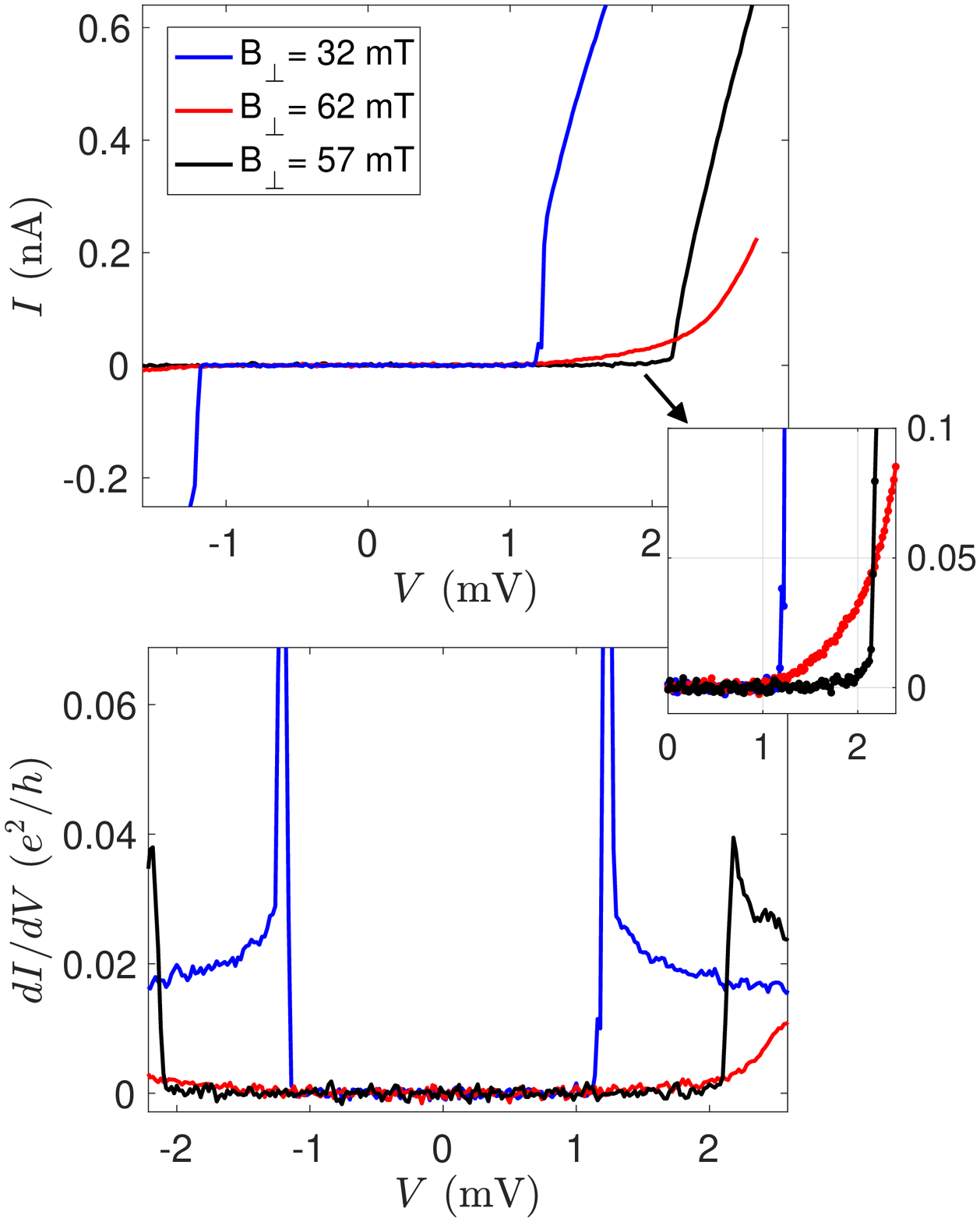} 
	\caption{Transport data $I$ \textit{vs.} $V$ (upper plot), and  $dI/dV$  \textit{vs.} $V$  (lower plot) for device B6 at selected values of perpendicular magnetic field: below the parity field, $B_\perp$\,=\,32\,mT (blue), at the parity field, $B^*_\perp=57$ mT (black), and above the parity field, $B_\perp=62$\,mT (red).  Although $V_c$ is approximately equal for blue and red data (inset), the transport characteristics are found to be substantially different. The conductance above $V_c$ decreases with $V$ for fields below $B^*$, as expected for Cooper-pair dominated transport, and increases with $V$ for $B_{||}$ above $B^*$, indicative of single-electron transport. Similar behavior is observed for perpendicularly-applied fields}	 
\end{figure}

\section{Current-voltage characteristics for varying magnetic fields}

In Figure 2 of the main text, we presented detailed current-voltage characteristics (IVC's) for device C5, measured in parallel magnetic fields, at field values below, at, and above $B^*$. In Fig. S2 and Fig. S3 we show similar data for device C3 in parallel magnetic fields, and device B6 in perpendicular magnetic fields. From Table I, one sees that these devices have quite similar values for charging energy $E_C$, as well as Josephson energy, $E_\mathrm{J}$. Correspondingly, the transport data for these devices is remarkably similar, although the applied parallel fields for device C3 are approximately six times higher than for device B6 in perpendicular fields.

Clearly the IVC's for both parallel and perpendicular fields are remarkably different above and below their respective parity fields, $B*$. For fields lower than their respective parity fields (blue data), the conductance $dI/dV$ is peaked just above $\pm V_c$, and strictly decreasing with $|V|$ outside the voltage gap. This can be understood as follows: (coherent) Cooper-pair dominated transport occurs more readily for energy differences approaching zero, which single-electron tunnelling involves breaking Cooper pairs, which increases with voltage bias.

\section{Estimate of $B_0$ from $V_c(B)$}

We assume that the homogenous superconducting gap, $\Delta(B)$, is suppressed by an applied magnetic field as, $\Delta(B)=\Delta_0(1-B^2/B_0^2)$, where $B_0$ is the depairing parameter. Using the Ambegaokar-Baratoff relation, the field-dependent Josephson energy is given by $E_{J}(B) =\Delta(B) R_\mathrm{Q}/2{R_J}$. The single-junction Bloch bandwidth, $W$, is the width of the lowest energy band in quasicharge for a current-biased junction \cite{likharev-zorin}, which depends solely on $E_C$ and $E_\mathrm{J}$, and can readily be calculated numerically for arbitrary $E_C$ and $E_\mathrm{J}$. For $E_\mathrm{J} \gg E_C$,

\begin{equation}
 W = \frac{16}{\sqrt{2\pi}} \hbar\omega_p\, (2 g)^{1/4} e^{ -\sqrt{32 g}}, 
\end{equation}
where $g=E_\mathrm{J}/4E_C$, and $\hbar\omega_p=\sqrt{8E_\mathrm{J} E_C}$. In Ref. \cite{cedergren-etal-17}, it was shown that the scaled critical voltage, $v=V_c/N\hbar\omega_p$, was set by the scaled, single-junction Bloch bandwidth, $w=W/\hbar\omega_p$, as 
\begin{equation}
v = a w^\alpha,
\end{equation}
where $a$ is a prefactor, and the exponent $\alpha=4/3-2K$, where $K$ is the Luttinger parameter, $K=\Lambda^{-1}\sqrt{E_\mathrm{J}/8E_C}$, and $\Lambda$ is the screening length.

 One can use these relations, along with the measured $V_c(B)$, for magnetic fields $B<B^*$, to estimate the depairing parameter, $B_0$. This is shown in Figs. S2 and S3, for respectively, parallel and perpendicular applied fields. For the fitting we have used a two parameter fit (prefactor $a$ and depairing parameter $B_0$), with the experimentally determined $E_\mathrm{J}$ and $E_C$ for each device as fixed parameters. The extracted depairing parameters are listed in Table I and II above.

\section{Additional details}

Additional experimental details can be found in the Supplemental Material of Cedergren \textit{et al.}, Ref. \cite{cedergren-etal-17}.


\begin{figure}
	\includegraphics[ width=\columnwidth]{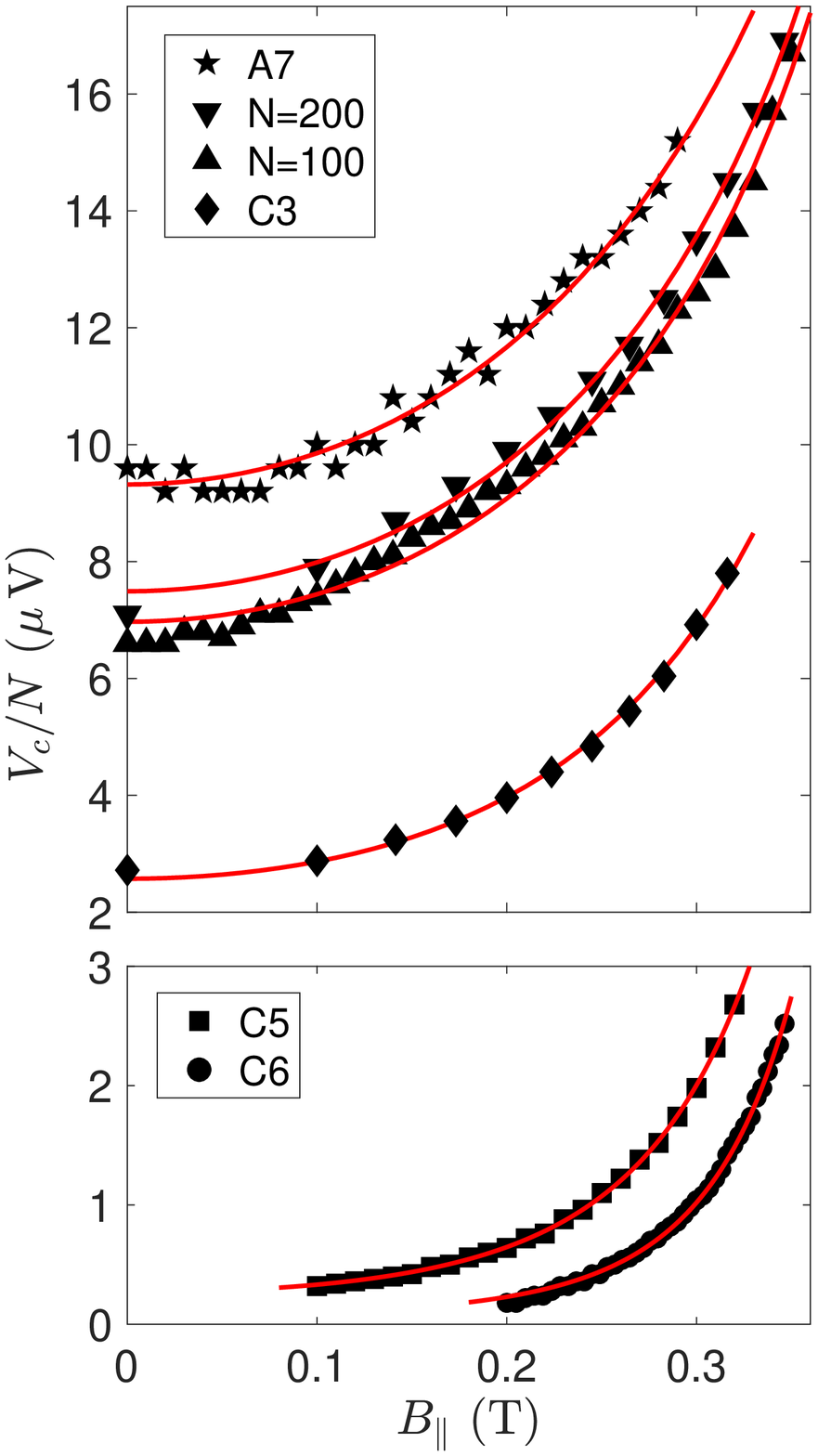} 
	\caption{Estimate of the depairing parameter, $B_0$, for devices measured in a parallel magnetic field by fitting to Eqn. (2) as described in the text. We have used two fit parameters, the prefactor $a$, and $B_0$, with $E_C$ and $E_\mathrm{J}$ experimentally determined from large scale IVC's as fixed parameters. See Table I for values of $B_0$ determined from the fits.}
	 
\end{figure}

\begin{figure}
	\includegraphics[ width=\columnwidth]{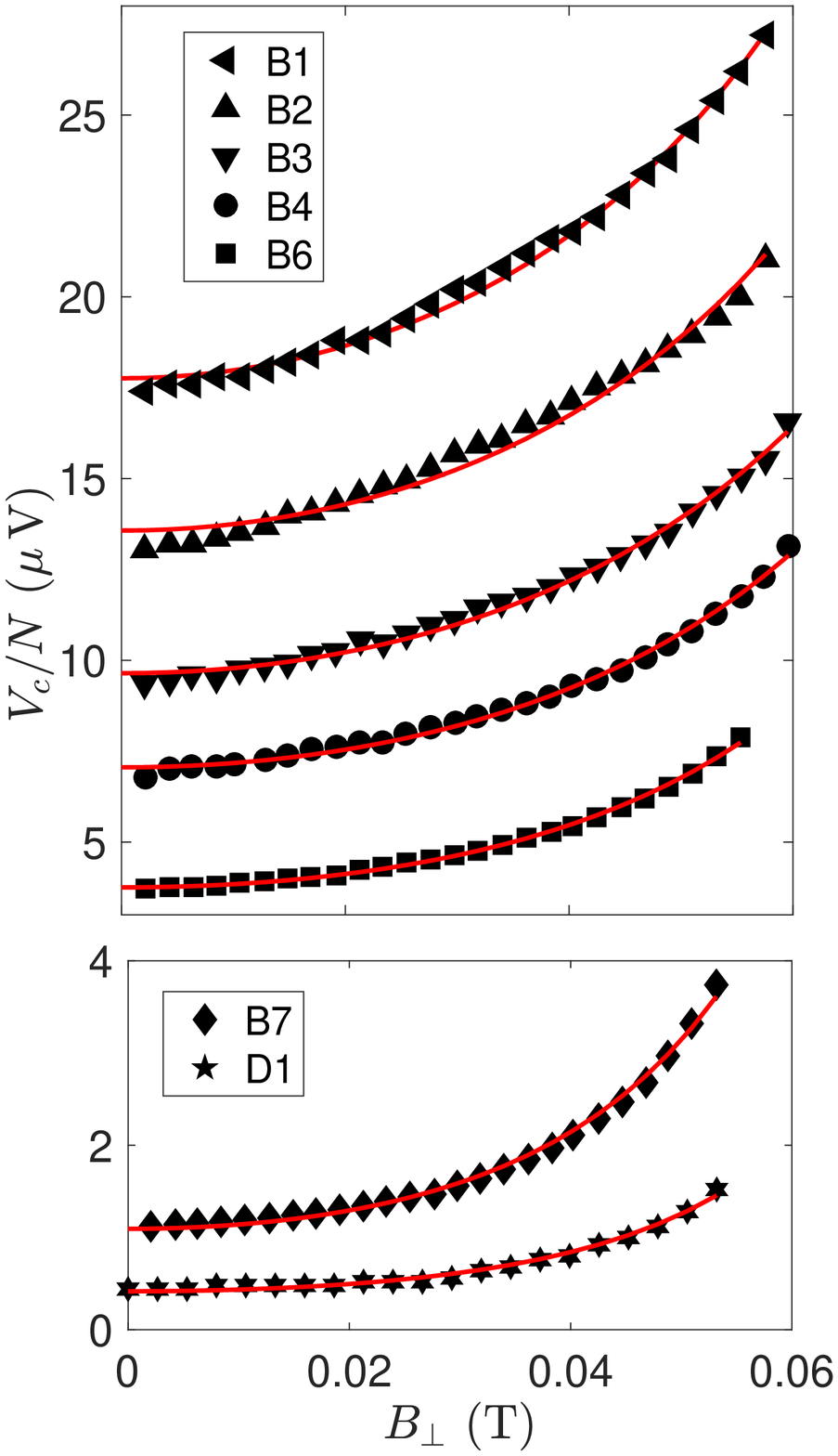} 
\caption{Estimate of the depairing parameter, $B_0$, for devices measured in a perpendicular magnetic field by fitting to Eqn. (2) as described in the text. We have used two fit parameters, the prefactor $a$, and $B_0$, with $E_C$ and $E_\mathrm{J}$ experimentally determined from large scale IVC's as fixed parameters. See Table I for values of $B_0$ determined from the fits.}
	 	 
\end{figure}

\end{thebibliography}

\end{document}